\documentclass[aps,amsmath,showpacs,amsfonts,10pt]{revtex4}
\usepackage{epsfig,graphicx}
\usepackage[english]{babel}
\usepackage{amsfonts}
\usepackage{amsmath}
\usepackage{latexsym}
\usepackage{graphics,bm}
\usepackage{natbib}
\usepackage{dcolumn}
\usepackage{bm}
\usepackage{rotating}

\begin{document}

\title{Detection of Weak Force using a Bose-Einstein Condensate}

\author{ Sonam Mahajan$^{1}$ , Tarun Kumar$^{1}$, Aranya B Bhattacherjee$^{2}$ and ManMohan$^{1}$}

\address{$^{1}$Department of Physics and Astrophysics, University of Delhi, Delhi-110007, India} \address{$^{2}$Department of Physics, ARSD College, University of Delhi (South Campus), New Delhi-110021, India}

\begin{abstract}
We investigate the possibility of detecting a weak coherent force by means of a hybrid optomechanical quantum device formed by a Bose Einstein Condensate (BEC) confined in a high quality factor optical cavity with an oscillatory end mirror. We show using the stochastic cooling technique that the atomic two-body interaction can be utilized to cool the mirror and achieve position squeezing essential for making sensitive measurements of weak forces. We further show that the atomic two-body interaction can also increase the signal to noise ratio (SNR) and decrease the noise of the off-resonant stationary spectral measurements.
\end{abstract}


\maketitle

\section{Introduction}

Recently, opto-mechanics is the subject of extensive theoretical and experimental investigations. The interaction between a movable mirror and the radiation field of an optical cavity provides a sensitive device which is able to detect weak forces. Significant examples of research in this area are the gravitational wave detection interferometers \citep{1,2} and atomic force microscopes \citep{3,4}. Over the past years, the field of laser cooling \citep{5,6,7} and gravitational wave detectors \citep{1,2} uses the interaction of mechanical and optical degrees of freedom via radiation pressure. In recent years, the outpouring interest in the application of radiation force is to change the center of mass motion of mechanical oscillators covering a vast range of scales from macroscopic mirrors in the Laser Interferometer Gravitational Wave Observatory (LIGO) project \citep{8} to nanomechanical cantilivers \citep{9,10,11,12,13,14}, vibrating micro-toroids \citep{15,16} and membranes \citep{17}.
The detection of displacement with high sensitivity is possible due to the capability of optical interferometry which have been shown recently in optical interferometry experiments\citep{18,19} and advances in gravitational wave detectors \citep{20}. Experimentally, highly sensitive optical displacement measurements can be done using a sensor which shows a direct effect of radiation pressure, back action effect that every optical experiment will be sensitive to if quantum noise is limited \citep{11}. The possibility of gravitational wave detection using atom field interferometers has been reported recently \citep{21}.
Much new possibility arises in cavity opto-mechanics when the experimental and theoretical tools of cavity quantum electrodynamics (QED) are combined with those of ultra cold gases \citep{22,23,24,25,treulein, szirmai,hunger2,chen, chiara, steinke, hunger3, chen2, zhang}. If a collection of atoms is placed in a high finesse optical cavity then the atoms collectively interact with the light mode thereby increasing the atom-field interaction. The coupling of the coherent motion of the condensate atoms trapped in an optical lattice, formed by a high finesse optical cavity and the intracavity field give rise to non-linear quantum optics \citep{25}. Even if the average photon number is as small as $0.05$, one can observe strong optical non-linearities \citep{25}. Earlier experiments have shown important progress in the field of cavity QED by combining it with ultracold atoms \citep{26,27,28}.
Thermal noise, which arises due to the mechanical motion of the mirror, is the major hurdle in calculating the sensitive optical measurements \citep{19,29}. It can be reduced using various feedback schemes based on the homodyne detection of the reflected light of the oscillator [30]. A continuous version of the stochastic cooling feedback technique, used in accelerators \citep{31}, helps in cooling the mirror of the opto-mechanical system as the feedback regularly "kicks" the mirror from the back and the position is regularly monitored using homodyne detection.
 In the presence of the feedback, it is important to describe the whole system quantum mechanically for two reasons. Firstly it permits to develop a condition for opto-mechanical systems under which the effects of quantum noise becomes visible and experimentally detectable. Lastly it establishes the ultimate limits of the proposed feedback scheme.
It is necessary to improve the signal to noise ratio or sensitivity of position measurements especially for gravitational wave detection \citep{1,2} or for meteorological applications \citep{32}. Detection of weak forces using cantilever is of much interest in many applications such as magnetometry of nanoscale magnetic particles \citep{rossel}, femtojoule calorimetry \citep{gimzewski} and other numerous types of force microscopy \citep{binning}. A measured force resolution of $5.6 \times 10^{-18} N/\sqrt{Hz}$ at $4.8 K$ in vacuum using cantilever based technology has been demonstrated \citep{stowe}. Motivated by these interesting developments in the field of ultracold atoms and cavity opto-mechanics, we propose in this paper a novel scheme to couple a Bose-Einstein condensate to an optical cavity with a movable mirror to detect weak forces using the stochastic cooling scheme.

\section{The Basic Model}

The system investigated here consists of a coherently driven Fabry-Perot cavity with one mirror ($M_{1}$) fixed while the second mirror ($M_{2}$) movable (Fig.1). This basic opto-mechanical setup is the basis of detecting weak forces in gravitational wave detectors \citep{1,2} and atomic force microscopy \citep{3,4}. In our model, we have in addition an elongated cigar shaped gas of $N$ two-level ultracold atoms of $^{87} Rb$ in the $|F=1>$ state having mass $m$ and transition frequency $\omega_{a}$ of the $|F=1>$ $\rightarrow$ $|F'=2>$ transition of the $D_{2}$ line of the $^{87} Rb$. The cloud of BEC is interacting with a single quantized mode of the cavity with frequency $\omega_{c}$. The cavity mode is also forming an optical lattice potential between the two mirrors. The sensitivity of the proposed quantum device to measure weak forces (shown in fig.1) is eventually determined by the quantum fluctuations. In order to minimize the quantum fluctuations associated with the various modes of the vibrating mirror, we consider the mirror as a single quantum-mechanical harmonic oscillator with frequency $\omega_{m}$ and mass $M$. Experimentally this approximation can be realized \citep{pinard}, if we use a bandpass filter in the detection loop, so that the frequencies are limited to a narrow bandwidth which includes a single mechanical resonance. An external pump with frequency $\omega_{p}$ is also incident from the fixed mirror, which is a constant source of photons for the cavity. We also consider that the optical cavity is of high quality factor since it ensures that light-field remains quantum-mechanical during the experiment. Intra-cavity photons exert pressure on the mirrors and this results in an opto-mechanical coupling between the cavity field and the movable mirror. The cavity light field exerts a force on the movable mirror which depends on the intracavity photons. The light in turn is phase shifted by an amount $2 \kappa l_{m}$, where $\kappa$ is the wave-vector and $l_{m}$ is the displacement of the mirror from its equilibrium position. In the adiabatic limit i.e $\omega_{m}<<c/2L$ ($L$ is the length of the optical cavity), single mode approximation of the cavity field is valid since photon scattering into other modes can be ignored. This system can be described by an opto-mechanical Hamiltonian in rotating wave and dipole approximation as \citep{22},

\begin{figure}[h]
\hspace{-0.0cm}
\includegraphics [scale=0.8]{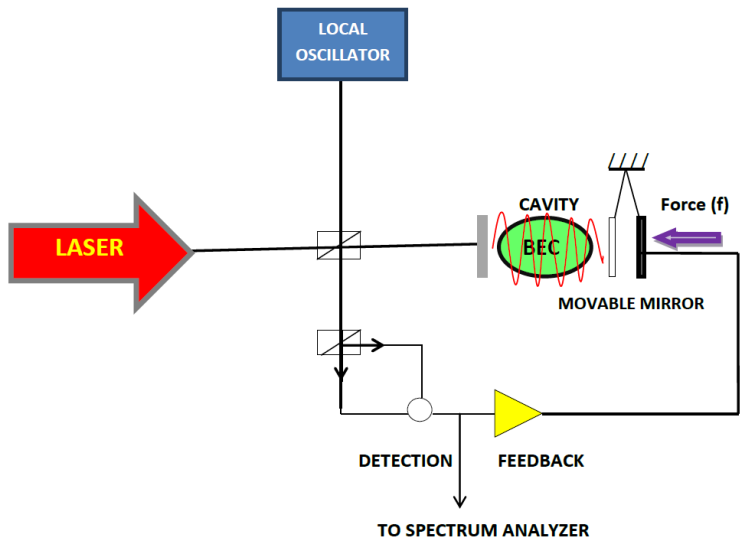}
\caption{Schematic representation of the setup. Figure shows an optomechanical system with Bose Einstein Condensate (BEC) confined in an optical cavity
with one mirror fixed and other is moving. Here the cavity mode is driven by the laser, which also provides the local oscillator for the homodyne measurement
 using the beam splitter. The external force (f) to be measured acts on the movable mirror.}
\label{1}
\end{figure}

\begin{eqnarray}\label{hom}
H &=& E_{0}\sum_{j}b_{j}^{\dagger}b_{j}+ J_{0}(\hbar U_{0}a^{\dagger}a+V_{cl}) \sum_{j}b_{j}^{\dagger}b_{j}+\dfrac{U}{2}\sum_{j}b_{j}^{\dagger}b_{j}^{\dagger}b_{j}b_{j} - \hbar \Delta_{c} a^{\dagger}a-i\hbar \eta (a-a^{\dagger})\nonumber \\ &+& \hbar \omega_{m} a_{m}^{\dagger}a_{m}-\hbar \epsilon \omega_{m} a^{\dagger}a(a_{m}+a_{m}^{\dagger})\;
\end{eqnarray}

where,

\begin{eqnarray}
U&=&\dfrac{4\pi a_{s}\hbar^{2}}{m}\int d^3 x|w(\vec{r})|^{4} , \nonumber \\
E_{0}&=&\int d^3 x w(\vec{r}-\vec{r}_{j})\left\lbrace \left( -\dfrac{\hbar^2 \nabla^{2}}{2m}\right) \right\rbrace w(\vec{r}-\vec{r}_{j}), \nonumber \\
J_{0}&=&\int d^3 x w(\vec{r}-\vec{r}_{j}) \cos^2(kx)w(\vec{r}-\vec{r}_{j}).
\end{eqnarray}

Here $b_{j}$, $a$ and $a_{m}$ are the condensate annihilation operators at the $j^{th}$ site, cavity mode annihilation operator and mirror mode (phonon) annihilation operator respectively. $\Delta_{c}=\omega_{p}-\omega_{c}$ is the cavity pump detuning. Here $\eta$ is the strength of the external pump, while $\omega_{m}$ is the frequency of oscillation of the movable mirror. The mirror-photon coupling is $G=\epsilon \omega_{m}$ and  $V_{cl}$ is the external classical potential. $U_{0}=\frac{g_{0}^{2}}{\Delta_{a}}$ is the optical lattice barrier height per photon where $g_{0}$ is the atom-photon coupling and $\Delta_{a}=\omega_{p}-\omega_{a}$ is the atom-pump detuning. We shall consider from now on $U_{0}>0$. In this case the condensate atoms are pulled towards nodes of the cavity light field and as a result the lowest bound state is localized at the nodes. This leads to a reduced coupling of the condensate atoms to the cavity field compared to that for $U_{0}<0$.  Also $E_{0}$ and $J_{0}$ are the effective onsite energies of the condensate defined in terms of the condensate atomic Wannier functions $w(\vec{r}-\vec{r}_{j})$. $U$ is the effective onsite atom-atom interaction energy, where $a_{s}$ is the $s$- wave scattering length. In deriving the above Hamiltonian Eqn.(\ref{hom}), we have ignored the tunneling of atoms into neighbouring wells. We can experimentally achieve this by tuning the optical lattice depth so that time scales over which tunneling takes place is much larger than the times scales over which the experiment is performed.  The cavity mode formed by the input pump laser couples to the mechanical oscillator (movable mirror) through radiation pressure and the condensate atoms through the dipole interaction. The back-action of the atoms and cantilever modifies the cavity field. The nonlinearity in Eqn. (\ref{hom}) arising from the coupling between the intracavity intensity and the position quadrature of the cantilever plays a significant role in the system dynamics. The system we are considering is basically an open system since the cavity field is damped due to the leakage of the photons through the cavity mirrors. Moreover the cantilever is connected to a bath at finite temperature. The mechanical oscillator is expected to undergo a pure Brownian motion driven by its contact with the thermal environment in the absence of any radiation pressure due to the cavity light field.
Let us now understand the origin of each of the terms appearing in the Hamiltonian of Eqn. (\ref{hom}). The first term is simply the onsite kinetic energy of the condensate. The second term is the coupling of the condensate with the cavity field which also includes the classical potential $V_{cl}$. The third term is the two-body atom-atom interaction term. The fourth term is the photon energy and the fifth term corresponds to the pump. The sixth term is the energy of the single mode of the mirror and the seventh term is the energy associated with the dispersive nonlinear coupling between the cavity field and the mirror.

Dissipation enters the system through its interaction with the external degrees of freedom. The cavity optical field is damped (decay constant $\gamma_{c}$) due to the leakage of photons through the mirrors. The mirrors couple the internal cavity modes with the external electromagnetic modes. The movable mirror is damped (decay constant $\gamma_{m}$) due to its interaction with the external modes. The condensate is more robust and there is no significant loss of atoms during the experimental time. The dynamics of the system can be described by the following set of coupled quantum Langevin equations (QLE),

\begin{equation}\label{a}
\dot{a}(t)=-i J_{0}U_{0}a(t) \sum_{j} b_{j}^{\dagger}(t) b_{j}(t)+i \Delta_{c} a(t)+i \epsilon \omega_{m} a(t) [a_{m}(t)+a_{m}^{\dagger}(t)]+\eta-\frac{\gamma_{c}}{2}a(t)+\sqrt{\gamma_{c}}a_{in}(t),
\end{equation}

\begin{equation}\label{b}
\dot{b_{j}}(t)=-i\frac{E_{0}}{\hbar}b_{j}(t)-i\frac{J_{0}}{\hbar}(\hbar U_{0}a^{\dagger}(t) a(t)+V_{cl})b_{j}(t)-i\frac{U}{\hbar}b_{j}^{\dagger}(t)b_{j}(t)b_{j}(t),
\end{equation}

\begin{equation}\label{am}
\dot{a_{m}}(t)=-i \omega_{m} a_{m}(t)+i \epsilon \omega_{m} a^{\dagger}(t) a(t)-\gamma_{m} a_{m}(t)+\sqrt{\gamma_{m}} \xi_{m}(t),
\end{equation}

where $a_{in}(t)$ and $\xi_{m}(t)$ are the input noise operators for the cavity field and mirror respectively with the following correlations \citep{giovanetti}:

\begin{equation}
\langle a_{in}(t) a_{in}(t')\rangle =\langle a_{in}^{\dagger}(t) a_{in}(t')\rangle =0,
\end{equation}

\begin{equation}
\langle a_{in}(t) a_{in}^{\dagger}(t)\rangle =\delta(t-t'),
\end{equation}

The steady state cavity field ($\beta$) can be derived from the Eqns.(\ref{a}, \ref{b}, \ref{am}) in terms of the steady state value of the phonon operator $\alpha$ and the number of atoms $N$ by putting the time derivative to zero.

\begin{equation}
\beta=\frac{\eta}{-i \Delta_{c}+\frac{\gamma_{c}}{2}+i J_{0}U_{0}N-i 2 Re(\alpha)\, \epsilon \, \omega_{m}},
\end{equation}

\begin{equation}
Re(\alpha)= \frac{\epsilon \, \omega_{m}^{2} |\beta|^{2}}{\omega_{m}^{2}+\gamma_{m}^{2}},
\end{equation}

 One can identify from the above equations that the steady state of the cavity field is influenced by the dynamics of the mirror and the atoms. The resonance frequency of the cavity is shifted due to its interaction with the mirror and the atoms in such a way so as to form a new stationary intensity. After a transient time, the cavity field changes depending upon the response of the field and the strength of the interaction with the condensate and the mirror.

We are now interested in the dynamics of fluctuations around the steady state. To this end, we linearize the QLE (Eqns. (\ref{a}, \ref{b}, \ref{am})) around the steady state as, $a(t)\rightarrow \beta+a(t)$, $a_{m}(t)\rightarrow \alpha+a_{m}(t)$ and $b_{j}(t)\rightarrow \frac{\sqrt{N}+\\b(t)}{\sqrt{M}}$. Here $\beta$, $\alpha$ and $b_{mf}=\sqrt{N/M}$ are the steady state of the photon, phonon and the atomic fields respectively. $M$ is the total number of lattice sites occupied by $N$ atoms. We also replace $b_{j}(t)$ by $b(t)$, assuming that all the sites of the optical lattice are identical. Consequently, we get,

\begin{equation}
\dot{a}(t)=\left [i \Delta_{d}-\frac{\gamma_{c}}{2} \right ]a(t)-i g_{c} \left [ b(t)+b^{\dagger}(t)\right ]+i G \beta \left [ a_{m}(t)+a_{m}^{\dagger}(t)\right ] +\sqrt{\gamma_{c}} a_{in}(t),
\end{equation}

\begin{equation}
\dot{b}(t)=-i \left [ \nu+2 U_{eff} \right ] b(t)-i U_{eff} b^{\dagger}(t)-i g_{c} \left [ a(t)+a^{\dagger}(t)\right ],
\end{equation}

\begin{equation}
\dot{a_{m}}(t)=-i \omega_{m} a_{m}(t)+i G \beta \left [ a(t)+a^{\dagger}(t)\right ]-\gamma_{m} a_{m}(t)+\sqrt{\gamma_{m}} \xi_{m}(t),
\end{equation}

where, $\Delta_{d}=-J_{0}U_{0}N+\Delta_{c}+2G \alpha$, $g_{c}=J_{0}U_{0} \beta \sqrt{N}$,  $G=\epsilon \omega_{m}$, $\nu=E_{0}/\hbar+J_{0}U_{0} \beta^{2}+J_{0}V_{cl}/\hbar$ and $U_{eff}= \frac{U N}{\hbar M}$. In the following, we will always take $\Delta_{d}=0$, relevant to many experimental situations.

We now introduce the following phase and amplitude quadratures: $X(t)=[a(t)+a^{\dagger}(t)]$, $Y(t)=i[a^{\dagger}(t)-a(t)]$, $Q(t)=[a_{m}(t)+a_{m}^{\dagger}(t)]$, $P(t)=i[a_{m}^{\dagger}(t)-a_{m}(t)]$, $Q_{c}(t)=[b(t)+b^{\dagger}(t)]$, $P_{c}(t)=i[b^{\dagger}(t)-b(t)]$, $X_{in}(t)=[a_{in}(t)+a_{in}^{\dagger}(t)]$, $Y_{in}(t)=i[a_{in}^{\dagger}(t)-a_{in}(t)]$ and $W(t)=i \sqrt{\gamma_{m}}[\xi_{m}^{\dagger}(t)-\xi_{m}(t)]$.

\begin{equation}\label{X}
\dot{X}(t)=-\frac{\gamma_{c}}{2}X(t)+\sqrt{\gamma_{c}}X_{in}(t),
\end{equation}

\begin{equation}
\dot{Y}(t)=-\frac{\gamma_{c}}{2}Y(t)-2 g_{c}Q_{c}(t)+2G \beta Q(t)+\sqrt{\gamma_{c}}Y_{in}(t),
\end{equation}

\begin{equation}
\dot{Q_{c}}(t)=(\nu+U_{eff})P_{c}(t),
\end{equation}

\begin{equation}
\dot{P_{c}}(t)=-(\nu+3 U_{eff})Q_{c}(t)-2 g_{c}X(t),
\end{equation}

\begin{equation}
\dot{Q}(t)=\omega_{m} P(t),
\end{equation}

\begin{equation}\label{P}
\dot{P}(t)=-\omega_{m} Q(t)+ 2 G \beta X(t)-\gamma_{m} P(t)+W(t).
\end{equation}

Here $W(t)=i \sqrt{\gamma_{m}}(\xi_{m}^{\dagger}(t)-\xi_{m}(t))$, satisfies the following correlation:

\begin{equation}
\langle W(t) W(t')\rangle = \frac{1}{2 \pi} \frac{\gamma_{m}}{\omega_{m}} [{f_{mr}(t-t')+i f_{mi}(t-t')}],
\end{equation}

where,

\begin{equation}
f_{mr}(t)=\int_{0}^{\overline{\omega}} d \omega \,  \omega \\ \cos \left ({\omega t} \right) \\ \coth \left ( {\frac{\hbar \omega}{2 k_{B} T}} \right ),
\end{equation}

\begin{equation}
f_{mi}(t)=- \int_{0}^{\overline{\omega}} d \omega \,  \omega \\ \sin \left ({\omega t} \right) .
\end{equation}

Here, $T$ is the bath temperature, $k_{B}$ is the Boltzmann constant and $\overline{\omega}$ the frequency cutoff of the reservoir spectrum. Note that the quantum Brownian motion of the mirror is non-Markovian in nature. Brownian noise is the thermal noise which arises due to the random motion of the movable mirror. The thermal noise term in the measured phase noise spectrum of the light reflected from the cavity is due to the quantum Brownian motion of the mirror \citep{30}. Here the antisymmetric part, corresponding to $f_{mi}$, is a direct consequence of the commutation relation and it is never a dirac delta while the symmetric part corresponds to $f_{mr}$ explicitly depends on temperature and becomes proportional to a Dirac Delta function only if the high temperature limit $k_{B}T>>\hbar \omega$ first and the infinite frequency cutoff limit $\omega \rightarrow \infty$, later are taken \citep{giovanetti}.
From the above Eqns.(\ref{X}-\ref{P}), we observe that, the phase quadrature of the cavity is only affected by the mirror position fluctuations $Q(t)$ and the condensate position fluctuations $Q_{c}(t)$. In general, we notice that it is only the phase quadratures which are affected by the fluctuations.

\section{Feedback in Stochastic cooling Scheme}

In most applications the mechanical oscillator (movable mirror) is used as a quantum meter to detect weak forces acting on it \citep{brag}. Consequently, the term that describes the action of the classical external force $f(t)$ on the mirror position $(a_{m}+a_{m}^{\dagger})$ is given as:

\begin{equation}
H_{f}=-\frac{\hbar}{2}(a_{m}+a_{m}^{\dagger})f(t).
\end{equation}

The force to be measured appears in the phase quadrature of the mirror only.

\begin{equation}\label{Pnew}
\dot{P}(t)=-\omega_{m} Q(t)+ 2 G \beta X(t)+f(t)-\gamma_{m} P(t)+W(t).
\end{equation}

Such a force can be measured by looking at the mirror's position quadrature $Q(t)$. In the large cavity bandwidth limit i.e $\gamma_{c}$ $>>$ $G \beta$,$\omega_{m}$, the cavity mode dynamics adiabatically follows that of the movable mirror. Therefore,

\begin{equation}
X(t)=\frac{2}{\sqrt{\gamma_{c}}} X_{in}(t),
\end{equation}

\begin{equation}
Y(t)=-\frac{4 g_{c}}{\gamma_{c}}Q_{c}(t)+\frac{4 G \beta}{\gamma_{c}}Q(t)+\frac{2}{\sqrt{\gamma_{c}}}Y_{in}(t).
\end{equation}

In the large cavity bandwidth limit, the weak force can be measured by monitoring the dynamics of the mirror position $Q(t)$, through the homodyne measurement of the phase quadrature $Y(t)$. Homodyne detection is a method of detecting frequency modulated radiation by non-linear mixing with radiation of a reference frequency (local oscillator). In homodyne detection , the reference frequency equals that of the input signal radiation. The dynamics of the cantilever can be controlled through a phase sensitive feedback loop which may be devised using the output of a homodyne measurement.
Interestingly, the phase quadrature $Y(t)$ now also depends on the position of the condensate $Q_{c}(t)$. It should then be possible to control $Y(t)$ through condensate parameters $\nu$ and $U_{eff}$. The experimentally detected quantity is the output homodyne photocurrent \citep{30},

\begin{equation}
Y_{out}=2 \eta' \sqrt{\gamma_{c}} Y(t)-\sqrt{\eta'} Y_{in}^{\eta'}(t),
\end{equation}

where, $\eta'$ is the detector efficiency, $Y_{in}^{\eta'}(t)$ is a generalized phase input noise, and $b_{\eta'}(t)$ is the generalized input noise, which satisfies the following correlations \citep{30}: $\langle b_{\eta'}(t)b_{\eta'}(t')\rangle $ $=$ $\langle b_{\eta'}^{\dagger}(t) b_{\eta'}(t')\rangle $ $=0$, $\langle b_{\eta'}(t)b_{\eta'}^{\dagger}(t')\rangle =\delta(t-t')$, $\langle b_{in}(t)b_{\eta'}^{\dagger}(t')\rangle  =   \langle b_{\eta'}(t)b_{in}^{\dagger}(t')\rangle  = \sqrt{\eta'} \delta(t-t')$. In the stochastic cooling scheme, the homodyne measurement provides a continuous monitoring of the oscillator's position and the feedback continuously kicks the mirror to put it back in its equilibrium position. When photons exert radiation pressure on the mirror in the cavity, it displaces the mirror from its equilibrium position.
To bring back the mirror to equilibrium position, a pressure is exerted on the mirror from opposite side through feedback. This helps in cooling down the mirror.
This technique uses the phase sensitive noise to cool the mirror. The feedback loop consists of a transducer which converts the random optical signal to a
stochastic electric signal which in turn mechanically drives the mirror's momentum.
This results in an additional term in the QLE for any generic operator $A(t)$ given by \citep{30},

\begin{equation}
\dot{A_{fb}}(t)=\frac{i \sqrt{\gamma_{c}}}{\eta'} Y_{out}(t-\tau) [g_{sc}P(t), A(t)],
\end{equation}

where, $\tau$ is the feedback loop delay time and $g_{sc}$ is a dimensionless feedback gain factor. In the limit of zero delay time $\tau\rightarrow 0$, we have the only non-zero dynamics of the feedback operator ,

\begin{equation}
\dot{Q_{fb}}=\frac{\sqrt{\gamma_{c}}}{\eta'}g_{sc} Y_{out}(t).
\end{equation}

As a result, the QLE for $Q(t)$ is modified as:

\begin{equation}
\dot{Q}(t)=\omega_{m} P(t)-8 g_{sc}g_{c} Q_{c}(t)+8 G \beta g_{sc} Q(t)+4 \sqrt{\gamma_{c}} g_{sc} Y_{in}(t)-\sqrt{\frac{\gamma_{c}}{\eta'}} g_{sc} Y_{in}^{\eta'}(t).
\end{equation}

The influence of the condensate on the mirror dynamics now appears due to the feedback loop.

\section{Stationary Oscillator Energy}

We now study the energy of the stationary state of the movable mirror, which is obtained in the $t \rightarrow \infty$ limit. We will particularly see how the two-body interactions $U_{eff}$ can be used to minimize the energy. The solutions of the QLE are generally obtained by Laplace transform numerically. Hence we obtain,

\begin{eqnarray}
Q(t)&=& H_{5}(t) Q(0)+4 \sqrt{\gamma_{c}} g_{sc} \int_{0}^{\infty} H_{5}(t') Y_{in}(t-t') dt'-g_{sc} \sqrt{\frac{\gamma_{c}}{\eta'}} \int_{0}^{\infty} H_{5}(t') Y_{in}^{\eta'}(t-t') dt'+ H_{1}(t) P(0) \nonumber \\
&+& H_{3}(t) Q_{c}(0)+H_{4}(t) P_{c}(0)+ \int_{0}^{\infty}[f(t-t')+W(t-t')]H_{1}(t')dt'+\int_{0}^{\infty} H_{2}(t')X_{in}(t-t')dt',
\end{eqnarray}

where,

\begin{equation}
H_{1}(t)=\mathfrak{L}^{-1} \left [ \frac{\omega_{m}[s^2+AB]}{H(s)} \right ]
\end{equation}

\begin{equation}
H_{2}(t)=\mathfrak{L}^{-1} \left [ \frac{ 4 G \beta \omega_{m}[s^2+AB]+32 A g_{sc} g_{c}^{2} (s+\gamma_{m})}{H(s)\sqrt{\gamma_{c}}} \right ]
\end{equation}

\begin{equation}
H_{3}(t)=\mathfrak{L}^{-1} \left [ \frac{- 8 g_{sc} g_{c} s (s+\gamma_{m})}{H(s)} \right ]
\end{equation}

\begin{equation}
H_{4}(t)=\mathfrak{L}^{-1} \left [ \frac{- 8 g_{sc} g_{c} A (s+\gamma_{m})}{H(s)} \right ]
\end{equation}

\begin{equation}
H_{5}(t)=\mathfrak{L}^{-1} \left [ \frac{(s+\gamma_{m})[s^2+AB]}{H(s)} \right ]
\end{equation}

\begin{equation}
H(s)=(s^2+AB)[\omega_{m}^{2}+(s+\gamma_{m})(s-8 G \beta g_{sc})],
\end{equation}

\begin{equation}
A=\nu+U_{eff},
\end{equation}

\begin{equation}
B=\nu+3 U_{eff}.
\end{equation}

\begin{figure}[h]
\hspace{-0.0cm}
\begin{tabular}{cc}
\includegraphics [scale=0.75]{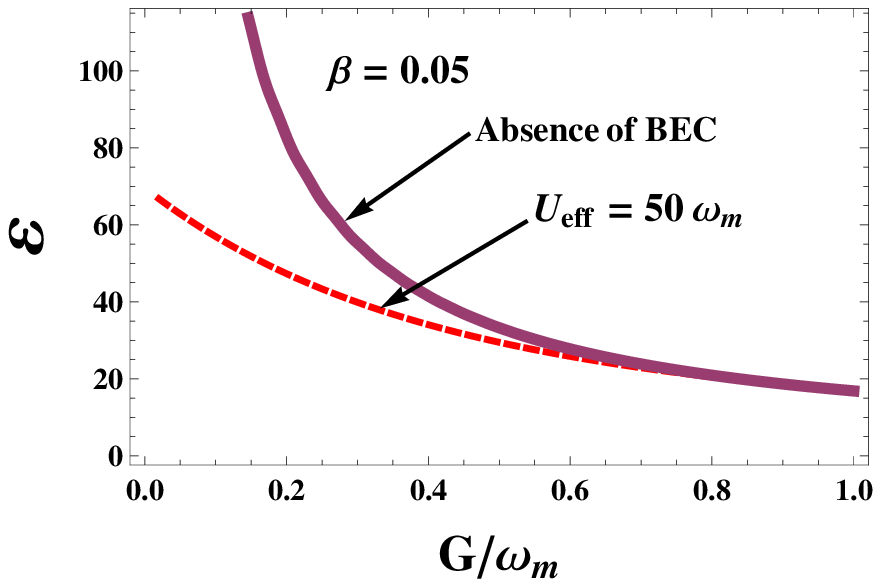}& \includegraphics [scale=0.75] {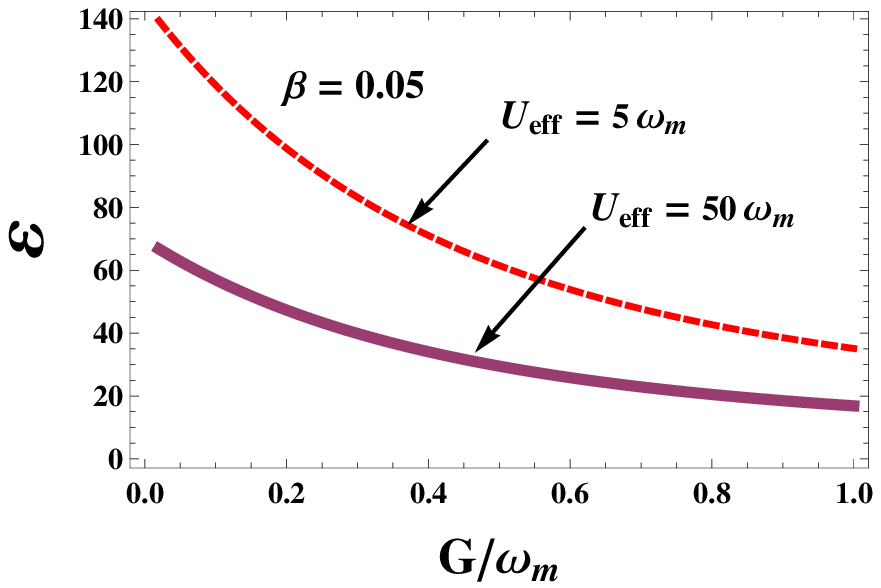}\\
 \end{tabular}
\caption{Plot of steady state energy (dimensionless with respect to $\omega_{m}$) as a function of mirror-photon coupling $G/\omega_{m}$.
The parameters used are: $\gamma_{m}/\omega_{m}=10^{-5}$, $\gamma_{c}/\omega_{m}=1$, $g_{sc}/\omega_{m}=-0.5$,$\eta'/\omega_{m}=0.8$,
 $\beta=0.05$,$k_{B}T/\hbar \omega_{m}=10^{5}$,$\nu/\omega_{m}=2$, $g_{c}=1.42 \omega_{m}$, $N=1000$.Here left hand plot
 shows the variation of the steady state energy in the absence of BEC and in the presence of BEC with $U_{eff}/\omega_{m}=50$.
Also right hand plot shows the variation of steady state energy for two values of atomic two-body interaction $U_{eff}/\omega_{m}=5$ (dashed line)
and $U_{eff}/\omega_{m}=50$ (solid line). }
\label{2}
\end{figure}

Similarly,

\begin{eqnarray}
P(t) &=& X_{7}(t) P(0)-\omega_{m} X_{6}(t) Q(0)+ X_{2}(t) P_{c}(0)+X_{3}(t) Q_{c}(0)+\int_{0}^{\infty}dt' X_{4}(t') X_{in}(t-t') \nonumber \\
 &-& 4 \omega_{m} \sqrt{\gamma_{c}} g_{sc} \int_{0}^{\infty} X_{6}(t') Y_{in} (t-t')dt' + \omega_{m} \sqrt{\frac{\gamma_{c}}{\eta'}} g_{sc} \int_{0}^{\infty} X_{6}(t') Y_{in}^{\eta'}(t-t') dt' \nonumber \\
 &+& \int_{0}^{\infty} X_{7}(t') [f(t-t')+W(t-t')] dt',
\end{eqnarray}

where,

\begin{equation}
X_{2}(t)=\mathfrak{L}^{-1} \left [ \frac{8 g_{sc} g_{c} A \omega_{m}}{H(s)} \right ]
\end{equation}

\begin{equation}
X_{3}(t)=\mathfrak{L}^{-1} \left [ \frac{8 g_{sc} g_{c} \omega_{m} s}{H(s)} \right ]
\end{equation}

\begin{equation}
X_{4}(t)=\mathfrak{L}^{-1} \left [ \frac{(4 G \beta (s^2+AB))(s-8G \beta g_{sc})-32 A \omega_{m} g_{sc} g_{c}^{2}}{H(s) \sqrt{\gamma_{c}}}  \right ]
\end{equation}

\begin{equation}
X_{6}(t)=\mathfrak{L}^{-1} \left [ \frac{s^2+AB}{H(s)} \right ]
\end{equation}

\begin{equation}
X_{7}(t)=\mathfrak{L}^{-1} \left [ \frac{(s-8 G \beta g_{sc})(s^2+AB)}{H(s)} \right ]
\end{equation}

\begin{figure}[h]
\hspace{-0.0cm}
\begin{tabular}{cc}
\includegraphics [scale=0.75]{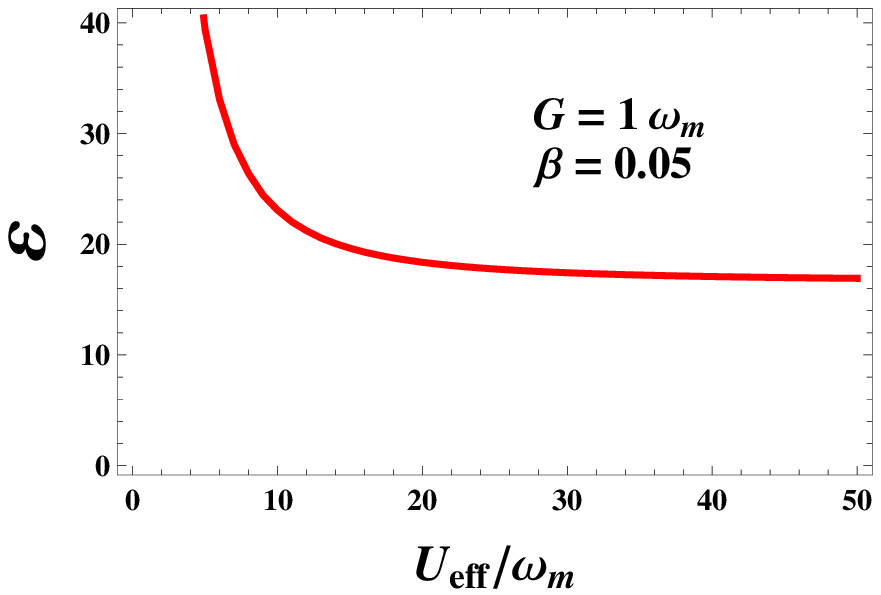}& \includegraphics [scale=0.75] {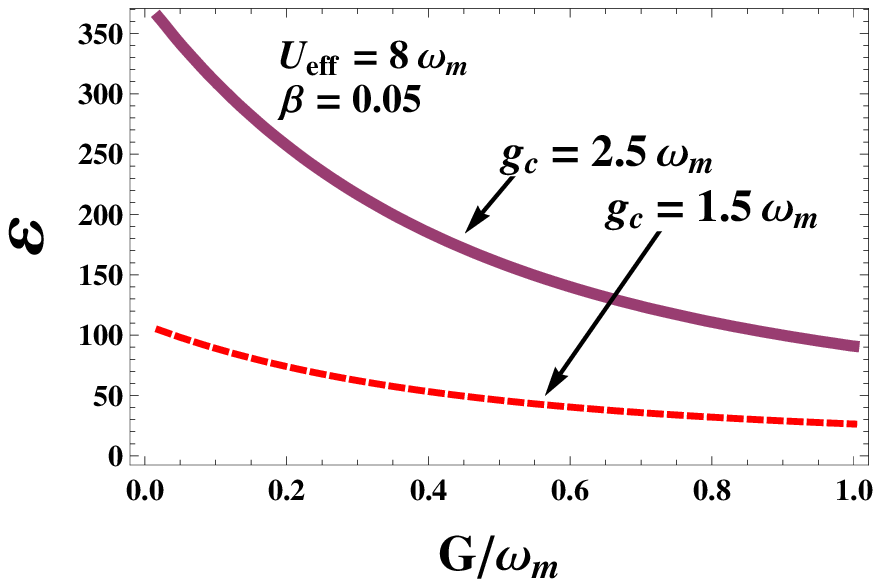}\\
\end{tabular}
\caption{Left plot: Plot of steady state energy as a function of $U_{eff}/\omega_{m}$. Right plot:  Plot of steady state energy versus mirror-photon coupling $G/\omega_{m}$ for two values of $g_{c}$. Parameters used are same as in Figure 2.}
\label{3}
\end{figure}

\begin{figure}[h]
\hspace{-0.0cm}
\includegraphics [scale=0.8]{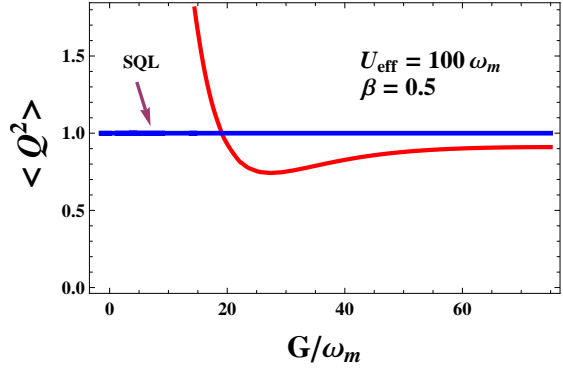}
\caption{Steady state position variance $\langle Q^{2}\rangle $ as a function of mirror-photon coupling $G/\omega_{m}$ for  $\gamma_{m}/\omega_{m}=10^{-4}$,
$\gamma_{c}/\omega_{m}=1.5$, $g_{sc}/\omega_{m}=-2.0$,$\eta'/\omega_{m}=1.0$,$g_{c}=14.23 \omega_{m}$,
 $\beta=0.5$,$k_{B}T/\hbar \omega_{m}=10^{4}$,$\nu/\omega_{m}=2$, $U_{0}/\omega_{m}=0.3$, $J_{0}/\omega_{m}=3$, $N=1000$ and $U_{eff}/\omega_{m}=100$.
 The horizontal full (blue line) line denotes the standard quantum limit (SQL).}
\label{4}
\end{figure}

Now using the following correlations: $\langle Y_{in}^{\eta'}(t)|Y_{in}^{\eta'}(t')\rangle =\delta (t-t')$, $\langle X_{in}(t)|X_{in}(t')\rangle =\delta (t-t')$, $\langle Y_{in}(t)|Y_{in}(t')\rangle = \delta(t-t')$, $\langle X_{in}(t)|Y_{in}(t')\rangle =i \delta (t-t')$, $\langle Y_{in}(t)|X_{in}(t')\rangle  = - i \delta(t-t')$, $\langle X_{in}(t)|Y_{in}^{\eta'}(t')\rangle =i \sqrt{\eta'} \delta(t-t')$, $\langle Y_{in}^{\eta'}(t)|X_{in}(t)\rangle =- i \sqrt{\eta'} \delta (t-t')$, $\langle Y_{in}(t)|Y_{in}^{\eta'}\rangle =\sqrt{\eta'} \delta(t-t')$ and $\langle Y_{in}^{\eta'}(t)|Y_{in}(t')\rangle  = \sqrt{\eta'} \delta(t-t')$, we find the stationary values of $\langle Q^{2}\rangle $ and $\langle P^{2}\rangle $,

\begin{equation}
\langle Q^{2}\rangle = g_{sc}^2 (8 \gamma_{c} + \frac{\gamma_{c}}{\eta'})\int_{0}^{\infty}[H_{5}(t)]^2 dt +\frac{\gamma_{m} k_{B}T}{\hbar \omega_{m}} \int_{0}^{\infty}[H_{1}(t)]^2 dt+\int_{0}^{\infty} [H_{2}(t)]^2 dt,
\end{equation}

\begin{equation}
\langle P^{2}\rangle = \omega_{m}^{2} g_{sc}^2 (8 \gamma_{c} + \frac{\gamma_{c}}{\eta'})\int_{0}^{\infty}[X_{6}(t)]^2 dt +\frac{\gamma_{m} k_{B}T}{\hbar \omega_{m}} \int_{0}^{\infty}[X_{7}(t)]^2 dt+\int_{0}^{\infty} [X_{4}(t)]^2 dt,
\end{equation}

The second term in the above equations for $\langle Q^{2}\rangle $ and $\langle P^{2}\rangle $ is the contribution of the quantum Brownian motion. We have used the high temperature approximation $Coth(\hbar \omega/2 k_{B} T)\approx 2k_{B}T/\hbar \omega$. The quality factor $Q_{f}=\omega_{m}/\gamma_{m}$ has to be high in order to reduce the affect of the thermal contribution.
We now study the stationary oscillator energy $U_{st}=\frac{\hbar \omega_{m}}{2}[\langle Q^2\rangle +\langle P^2\rangle ]$ and investigate the influence of the condensate two-body interaction $U_{eff}$ and the atom-photon coupling constant $g_{c}$ on the cooling of the mirror.
In the figures, we will always plot the dimensionless steady state energy $\epsilon= 2 U_{st}/\hbar \omega_{m}$. The left plot of Fig.2 illustrates the steady state energy as a function of dimensionless mirror-photon coupling $G$ in the absence of BEC (solid line) and in the presence of BEC (dashed line). Clearly we observe that the oscillator energy is reduced in the presence of BEC for small values of $G$. For higher values of $G$, the two plots merge. The right plot of Fig.2 depicts the steady state energy as a function of $G$ for two values of $U_{eff}=5 \omega_{m}$ (dashed line) and $U_{eff}=50 \omega_{m}$ (solid line). A substantial lowering of oscillator energy is noticed for higher $U_{eff}$. Variation of the steady state oscillator energy with $U_{eff}$ is shown in the left plot of Fig.3. A rapid initial decline in the oscillator energy with increasing $U_{eff}$, followed by a steady value of the energy for higher $U_{eff}$ is noticed.

Coupling between the condensate atoms, the cavity field and the mirror mode leads to a resonant energy exchange between the three systems. Such energy exchange leads to normal mode splitting \citep{22}. In our earlier work \citep{22}, we had shown that the condensate atoms participate in the energy exchange only in the presence of a finite two-body interaction $U_{eff}$ indicating that the Bogoluibov mode of the condensate is involved in the three mode coupling. In a recent experiment \citep{24}, coupling of a cloud of ultracold atoms to an optical resonator suggest that the Bogoluibov modes interacting significantly with the cavity field are those with momentum $\pm 2 k_{c}$ ($k_{c}$ is the cavity wave-number). The observed decrease in the energy of the mirror with increase in $U_{eff}$ could be due to resonant transfer of energy from the mirror mode to the condensate mode via the cavity mode. The indirect coupling (mediated by the cavity field) to the collective excitations of the condensate determines the number of thermal excitations in the mechanical mode and hence its energy. The BEC can absorb energy taken from the mirror by the cavity field. Increase in membrane (coupled to a BEC via cavity mode) dissipation due to increase in atom number has been demonstrated experimentally \citep{camerer}. Note that $U_{eff}$ is directly proportional to the atom number $N$. This experiment confirms our result that increase in $U_{eff}$ decreases the mirror energy.

The influence of the three mode coupling is also seen in right plot of Fig.3 where we have plotted the steady state energy of the oscillator as a function of $G$ for two values of atom-photon coupling $g_{c}$. As $g_{c}$ increases, the energy of the mechanical mode increases, suggesting that energy is being transferred from the condensate mode to the mirror mode via the cavity mode due to atomic back action. This observation is consistent with earlier results \citep{paternostro}. The above results reveal that on one hand increasing $U_{eff}$ decreases the energy of the mirror, on the other hand increasing $g_{c}$ increases the energy of the mirror. In order to measure any weak force acting on the mirror an optimal value of mirror energy can be achieved with the two handles, $U_{eff}$ and $g_{c}$.

 This stochastic cooling scheme can also be used to achieve steady state position squeezing, i.e, to overcome the standard quantum limit $\langle Q^{2}\rangle _{st}<1$. The possibility to beat the standard quantum limit for the oscillator position uncertainity is shown in Fig.4, where $\langle Q^{2}\rangle _{st}$ is plotted as a function of $G$. The standard quantum limit is seen to be beaten in some specific parameter regime as found in Fig.4.

\section{Noise Power Spectrum}

\begin{figure}[h]
\hspace{-0.0cm}
\begin{tabular}{cc}
\includegraphics [scale=0.82]{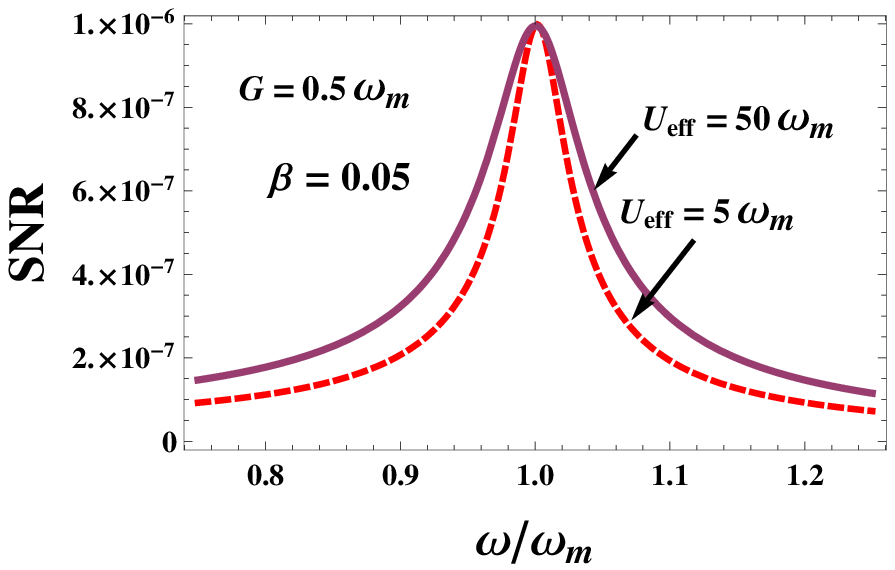}& \includegraphics [scale=0.75] {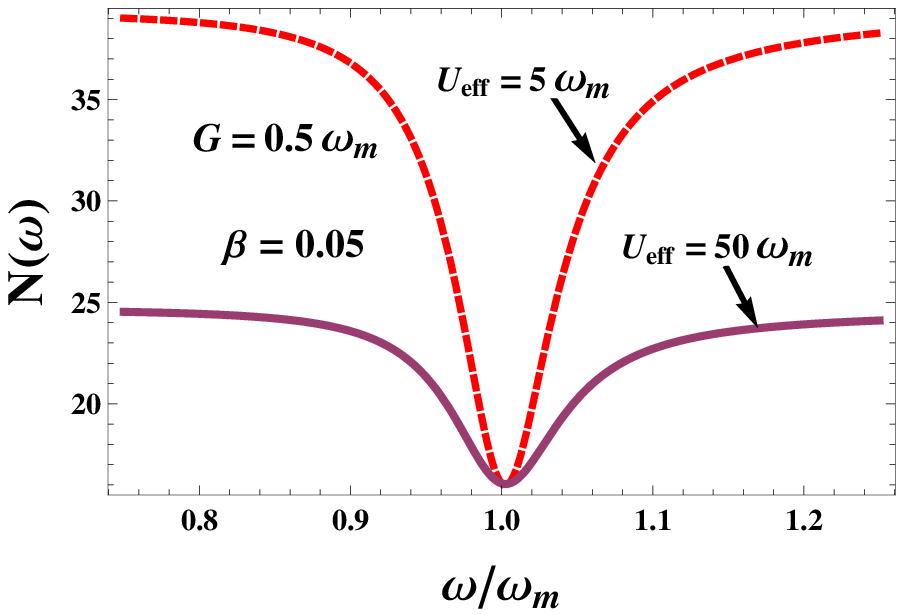}\\
 \end{tabular}
\caption{Plot of stationary signal to noise ratio as a function of $\omega/\omega_{m}$ in the case of ideal impulsive force ($f(\omega)=$ constant).
The parameters used are: $\gamma_{m}/\omega_{m}=10^{-5}$, $\gamma_{c}/\omega_{m}=1$, $g_{sc}/\omega_{m}=-0.5$, $\eta'/\omega_{m}=0.8$,
 $\beta=0.05$, $k_{B}T/\hbar \omega_{m}=10^{5}$, $\nu/\omega_{m}=2$, $g_{c}=1.42 \omega_{m}$ ,$N=1000$, $f(\omega)=10^{-5}\omega_{m}$,
 $T_{m}=10^{6}/\omega_{m}$, $G/\omega_{m}=0.5$. Here left hand plot shows the stationary signal to noise ratio plot for the
 two values of atomic two-body interaction  $U_{eff}/\omega_{m}=5$ (dashed line) and $U_{eff}/\omega_{m}=50$ (solid line). The right plot shows the noise spectrum for the same parameters as for the left plot.}
\label{5}
\end{figure}

The oscillator energy studied in the previous section, ultimately has to be measured in the form of the noise power spectrum of the mirror. Here we investigate the signal to noise ratio (SNR) of the optomechanical device. The signal corresponding to the spectral measurement in terms of the directly measured quantity "output homodyne photocurrent" $Y_{out}(t)$ is defined as \cite{30}

\begin{equation}
S(\omega)= |\int_{-\infty}^{\infty} dt e^{-i \omega t} \langle Y_{out}(t) F_{T_{m}}(t)\rangle |,
\end{equation}

where,

\begin{equation}
\langle Y_{out}\rangle = \frac{-8 \eta' g_{c}}{\sqrt{\gamma_{c}}}\langle Q_{c}\rangle +\frac{8 \eta' G \beta}{\sqrt{\gamma_{c}}}\langle Q\rangle .
\end{equation}

Here $F_{T_{m}}(t)$ is a filter function approximately equal to $1$ in the time interval $[0,T_{m}]$, in which the spectral measurement is performed and equal to zero otherwise \citep{30}. For stationary spectral measurements, the measurement time $T_{m}$ is taken to be much larger than the oscillator relaxation time $1/\gamma_{m}$ i.e. $T_{m}>> 1/\gamma_{m}$. In this limit, for very large measurement time $T_{m}$, one has $F_{T_{m}} \approx 1$. In this case, the oscillator is relaxed to equilibrium. This yields,

\begin{equation}
S(\omega)=\frac{8 \eta' G \beta}{\sqrt{\gamma_{c}}}|H_{1}(\omega) f(\omega)|,
\end{equation}

where, $f(\omega)$ is the Fourier transform of the force while  $H_{1}(\omega)$ is the Fourier transform of  $H_{1}(t)$ . Note that the signal is independent of the BEC. The noise corresponding to the signal is given by its variance as \citep{30},

\begin{equation}
N^{2}(\omega)= \int_{-\infty}^{\infty} dt F_{T_{m}}(t) \int_{-\infty}^{\infty} dt' F_{T_{m}}(t') e^{-i \omega (t-t')} \langle Y_{out}(t)|Y_{out}(t')\rangle _{f=0}.
\end{equation}

Making use of the previously derived results, we arrive at the final form of the noise spectrum,

\begin{eqnarray}
N^{2}(\omega)&=& \frac{64 \eta'^{2} g_{c}^{2}}{\gamma_{c}}T_{m} |Q_{N}(\omega)|^{2} - \frac{128 \eta'^{2} G \beta g_{c} T_{m}}{\gamma_{c}} |H_{2}(\omega)Q_{N}(\omega)|  \nonumber \\
&+& \frac{64 \eta'^{2} G^{2} \beta^{2}}{\gamma_{c}}T_{m} \left [ g_{sc}^{2}(8 \gamma_{c} +\frac{\gamma_{c}}{\eta'})|H_{5}(\omega)|^{2}+\frac{\gamma_{m} k_{B} T}{\hbar \omega_{m}}|H_{1}(\omega)|^{2}+|H_{2}(\omega)|^{2} \right ] \nonumber \\
&+& 8 \eta'^{2} T_{m}+ \eta' T_{m} + G \beta g_{sc} \left [ 128 \eta'^{2} +16 \eta' \right ] T_{m} |H_{5}(\omega)|,
\end{eqnarray}

where,

\begin{equation}
Q_{N}(t)=\mathfrak{L}^{-1} \left [ \frac{-4 A g_{c} \omega_{m}^{2}-4 A g_{c}(s+\gamma_{m})(s-8 G \beta g_{sc})}{\sqrt{\gamma_{c}}H(s)} \right ]
\end{equation}

The influence of BEC appears in the noise. The left plot of Fig.5 displays the SNR for two values of the atomic two-body interaction $U_{eff}$. No change in the SNR signal is seen at resonance. The main effect of the parameter $U_{eff}$ on the spectrum is the modification of the susceptibility due to the increase in the mechanical damping, which is responsible for the broadening of the spectrum. As seen above, the BEC does not influence the signal but only modifies the noise (Eqn.49). The right plot of fig.5 shows the influence of the BEC on the noise. Interestingly increasing $U_{eff}$ suppresses the off-resonant noise but not the resonant noise. This implies that a off-resonant measurement would be suitable for the measurement of the weak force acting on the mirror. Other schemes such as non-stationary measurements could be also suitable \citep{giovanetti99}.
We now discuss the experimental feasibility of the relevant parameters. For a BEC containing an order of $10^5$ $^{87}Rb$ atoms interacting with light field of an optical ultra high finesse Fabry Perot cavity, the strength of the coherent coupling is $g_{0} = 2 \pi  \times 10.9$ MHz \citep{24}, ($2 \pi \times 14.4 MHz$ \citep{25}) which is more than the decay rate of the intracavity field $\gamma_{c} = 2\pi \times 1.3$ MHz \citep{24} ($2 \pi \times 0.66$ MHz \citep{25}). The contribution of the kinetic and potential energy is about $\nu = 35$ kHz \citep{24} ($\nu = 49$ kHz \citep{25}). Also the coherent amplification or the damping of atomic motion is neglected as the temperature ($T_{c}$) of the condensate gas $T_{c} <<  \hbar \gamma_{c} /k_{B}$. The atom pump detuning is $2 \pi \times 32$ GHz. The mechanical frequency $\omega_{m} = 2 \pi \times 73.5$ MHz and damping rate of mirror $\gamma_{m} = 2 \pi \times 1.3$ kHz is recently reported in \citep{schliesser08}. The coupling rate is $G = 2 \pi \times 2$ MHz. The atom field coupling is reduced as there is decrease in the energy of the cavity mode due to the loss of photons through the cavity mirrors. By using high quality factor cavities, this loss of photons can be minimized. For the perfect homodyne detection, the detector efficiency is $\eta' = 1$ and if the additional noise is taken into account then due to the inefficient detection i.e. for the general case , $\eta' < 1$ \citep{giovanetti99}. Also in typical optomechanical experiments \citep{19,29,cohadon,pinard1}, the limit  $\gamma_{m}$ $<<$ $\omega_{m}$ $<<$ $k_{B}T$ is always taken.

\section{Conclusions}
In conclusion we have studied how stochastic cooling scheme can improve the sensitivity of an optomechanical device consisting of a gas of ultra-cold atoms confined in an optical cavity with a movable end mirror. We have seen that the atom-atom two body interaction can effectively cool the mechanical mirror and also achieve steady state position squeezing by beating the standard quantum limit. The atom-photon interaction on the other hand heats up the mirror. The feedback loop introduces the condensate as a new handle to control the mirror dynamics. We have also analyzed the sensitivity (SNR) of the optomechanical quantum device for the case of stationary position spectral measurements for the detection of weak forces. We found that the presence of the condensate does not change the SNR at resonance. However if we increase the atomic two-body interaction, the off-resonant SNR signal broadens and the off-resonant SNR increases. Also the two-body interaction significantly reduces the noise of the spectral measurements. The system presented here appears as novel optomechanical quantum device to measure weak forces. A coherent control of this device can be achieved through the two-body atom-atom interaction which can be manipulated either by the number of atoms or the $s$ wave scattering length.

\section{Acknowledgements}
 A. Bhattacherjee acknowledges financial support from the Department of Science and Technology, New Delhi for financial assistance vide grant SR/S2/LOP-0034/2010. Sonam Mahajan acknowledges University of Delhi for the University Teaching Assistantship.

\end{document}